\documentclass[%
reprint,
preprintnumbers,
 amsmath,amssymb,
 graphicx,
aps,
prb,
]{revtex4-1}

\usepackage{graphicx}
\usepackage{dcolumn}
\usepackage{bm}
\usepackage[mathlines]{lineno}

\begin{document}
\preprint{Journal of the Physical Society of Japan}

\title{Anharmonic Decay of Coherent Optical Phonons in Antimony}

\author{Muneaki Hase$^{1,2}$, Kiminori Ushida$^3$, and Masahiro Kitajima$^{1,2,4}$}
\email{mhase@bk.tsukuba.ac.jp}
\affiliation{$^1$Division of Applied Physics, Faculty of Pure and Applied Sciences, University of Tsukuba, 1-1-1 Tennodai, Tsukuba 305-8573, Japan \\
$^2$National Institute for Materials Science, 1-2-1 Sengen, Tsukuba, 305-0047, Japan \\
$^3$Department of Chemistry, School of Science, Kitasato University, 1-15-1 Kitasato, Minami-ku, Sagamihara, Kanagawa 252-0373, Japan \\
$^4$LxRay Co., Ltd, 1-22 Tazawa bldg., Koshien-2Bancho, Nishinomiya 663-8172, Japan 
}

\begin{abstract}
Anharmonic decay of coherent optical phonons in semimetal Sb has been investigated by using a femtosecond pump-probe technique. The coherent $A_{1g}$ mode is observed in time 
domain in a wide temperature range of 7 - 290 K. The decay rate (the inverse of the dephasing time) systematically increases as the lattice temperature increases, which is well explained 
by anharmonic phonon-phonon coupling, causing decay of the optical phonon into two acoustic phonon modes. 
The frequency of the $A_{1g}$ mode decreases with the temperature, which is interpreted to the results of both thermal expansion and anharmonic phonon-phonon coupling.
The temperature dependence of the amplitude of the coherent $A_{1g}$ 
mode exhibits a decrease with the lattice temperature, which is well reproduced by considering the peaked intensity of spontaneous Raman scattering assuming a Lorentzian line shape 
with the linewidth controlled by the anharmonic decay, and this model can be applicable to other metallic system, like Zn. 
\end{abstract}

\date{\today}


\maketitle

\section{Introduction}

In recent years, coherent phonons have been studied by using a femtosecond pump-probe 
technique in various material systems, including semiconductors,\cite{Cho90,Lee02} metals,\cite{Cheng90,Hase05} superconductors,\cite{Misochko00,Takahashi11,Kim12} organic conductors,\cite{Onda08} and phase change materials.\cite{Forst00,Hase09} 
The coherent phonons can be generated by ultrashort laser pulses with a high degree of temporal coherence. The nature of the coherent phonons has 
been extensively studied in semimetals\cite{Garrett96} and semiconductors,\cite{Kutt92} where the main focus was the generation mechanism. 
Because the coherent phonons are lattice vibrations having the same phase in time, dephasing of the coherent phonon is directly monitored by optical pump-probe measurements.  

The studies on the dephasing process are especially important for understanding the nature of in-phase coherent phonons. 
The dephasing process of coherent phonons excited by picosecond pulses has been examined using time-resolved CARS (coherent anti-Stokes Raman scattering).\cite{Linde80,Bron86} 
The phonon decay rate can be generally described as the sum of the anharmonic decay rate and the pure dephasing rate.\cite{Laubereau78} 
The main channels for the relaxation of incoherent optical phonons in semiconductors are thought to be the dephasing originating from the phonon-phonon interaction caused by anharmonicity 
of the lattice potential. In this channel, the excited optical phonons decay into acoustic phonons.\cite{Klemens66,Menendez84} A direct comparison has been made between the results obtained by time-resolved 
CARS and Raman scattering spectroscopies.\cite{Laubereau78,Bron86} 
Anharmonic phonon decay of coherent phonons was first confirmed by Hase {\it et al.} in Bi by measuring the temperature dependence of the dephasing time or decay rate (the inverse of the dephasing time).\cite{Hase98} 
It has been also shown that the dephasing of the coherent optical phonon is very sensitive to the density of lattice defect (vacancy).\cite{Hase00,Hase10} 
However, systematic study of the dephasing of coherent optical phonons in semimetals other than Bi excited by femtosecond laser pulses is still few. 

In this paper, dephasing process of coherent optical phonons in semimetal antimony (Sb) is studied by femtosecond pump-probe reflectivity measurements in a temperature range 
from 7 to 290 K. The amplitude, the dephasing time, and the frequency of the coherent optical phonons have precisely been measured. The decay rate and frequency show the systematic change as a function of the lattice temperature, similar to those in Bi, and reproduced by the model based on anharmonic lattice effects. The amplitude of the coherent phonon also exhibits systematic decrease with increasing the lattice temperature, which is well fit by the model based on the phonon occupation number proposed by Misochko {\it et al}.\cite{Misochko06} 

\section{Experimental technique}

The samples used in this study was a single crystal of Sb with cut and polished with the (0001) surface. The femtosecond pump-probe measurements were carried
out in a temperature range from $\approx$7 to 290 K using a closed-cycle cryostat. The light source used was a mode locked Ti:sapphire laser with a central wavelength of 800 nm, providing $\approx$20 fs pulses at 
the repetition rate of 87 MHz. 
The pump and probe beams were polarized orthogonal each other to avoid the scattered pump beam. Both pump- and probe-beams were focused onto a diameter of $\approx$100 $\mu$m on the sample. 
The average power of the pump and probe beams were fixed at 120 mW and 5 mW, 
respectively, from which we estimated the pump fluence to be 18.4 $\mu$J/cm$^{2}$ at 120 mW. By changing the optical path length of the probe beam, the reflectivity change ($\Delta R/R$) was recorded as a function of the delay time. 
This isotropic reflectivity measurement enable us to dominantly detect the fully symmetric $A_{1g}$ mode in Sb, while the non-symmetric $E_{g}$ mode is generally masked.\cite{Cheng90} Thus the determination of the phonon parameters, i.e., the amplitude, the dephasing time, and the frequency is accurate in the present study because a single damped harmonic oscillator model can fit the coherent phonon signal with smallest number of parameters.\cite{Hase98} 

\section{Experimental results and analysis}
Figure 1 shows the transient reflectivity change ($\Delta R/R$) recorded in Sb single crystal at the lattice temperatures between 7 and 290 K. 
The coherent oscillation of the fully symmetric $A_{1g}$ mode ($\approx$ 150 cm$^{-1}$ = 4.52 THz at 290 K),\cite{Cheng90,Ishioka08} superimposed on the photoexcited carrier background, is observed in the wide temperature range. 
As the temperature is lowered, the dephasing time becomes longer and the amplitude of the coherent $A_{1g}$ mode increases. The frequency, the amplitude, and the dephasing time of the coherent $A_{1g}$ 
mode at various temperatures were determined in the present study by fitting the time-domain data to a damped harmonic oscillation with a exponential decay function:\cite{Hase98}
\begin{widetext}
\begin{eqnarray}
\frac{\Delta R (t)}{R} = A\exp\Bigl(- \frac{t}{\tau}\Bigr)\cos(\omega_{A_{1g}} t + \phi) + B \Bigl[\exp\Bigl(- \frac{t}{\tau_{1}}\Bigr) - \exp\Bigl(- \frac{t}{\tau_{2}}\Bigr)\Bigr],
\end{eqnarray}
\end{widetext}
where $A$ is the amplitude, $\omega_{A_{1g}}$ is the frequency, $\tau$ is the dephasing time, and $\phi$ is the initial phase of the coherent $A_{1g}$ mode. The second term arises from the 
photoexcited carriers. Here $B$ is the amplitude, and $\tau_{1}$ and $\tau_{2}$ are the relaxation time, and the rising time of the electric component, respectively. 
The fit results in Fig. 1 are satisfactory good and thus we obtain the phonon parameters. The initial phase ($\phi$) obtained is nearly zero, meaning the coherent oscillation is cosine-like. 
This phase value matches with the 
previous results in Sb that the generation mechanism of the coherent $A_{1g}$ mode is governed by the displacive excitation of coherent phonon (DECP) model\cite{Cheng90,Zeiger92} or Raman scattering finite lifetime (RSFL) model,\cite{Riffe07,Ishioka08} the latter of which took both the finite lifetime of the coupled charge carrier density (a nature of DECP) and stimulated Raman scattering into account. 
The amplitude of the electric component, $B$, exhibits increase by $\sim$ 15 \% as the temperature is lowered as seen in Fig. 1, suggesting the photoexcited carrier density increases with decreasing the lattice temperature.\cite{Hase05}  
We will discuss the temperature dependence of the parameter $B$ together with the amplitude of the coherent $A_{1g}$ mode ($A$) in the later sections. The values of the relaxation time, $\tau_{1}$ and $\tau_{2}$, do not systematically change with the temperature, and therefore we would not discuss the time constants of the electric component in detail in the present paper. 
Hereafter, the temperature dependences of the decay rate, the frequency, and the amplitude of the coherent optical phonons are mainly focused and explored. 
\begin{figure}
\includegraphics[width=8.0cm]{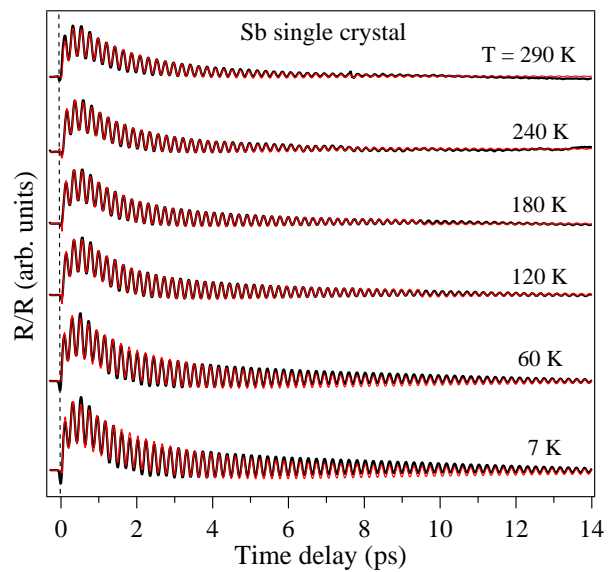}
\caption{The time-resolved reflectivity changes for Sb single crystal at temperatures between 7 K and 290 K. The black thin lines are the data, while the red lines are the fit with Eq. (1). 
}
\label{Fig1}
\end{figure}

In Fig. 2 the decay rate (the inverse of the dephasing time) of the $A_{1g}$ mode at different temperatures obtained by the time-domain measurement are plotted. 
The frequency of the coherent $A_{1g}$ mode is also plotted in the inset. The frequency of the $A_{1g}$ mode shifts from 4.67 THz to 4.52 THz when the temperature 
increases from 7 to 290 K. 
\begin{figure}
\includegraphics[width=7.5cm]{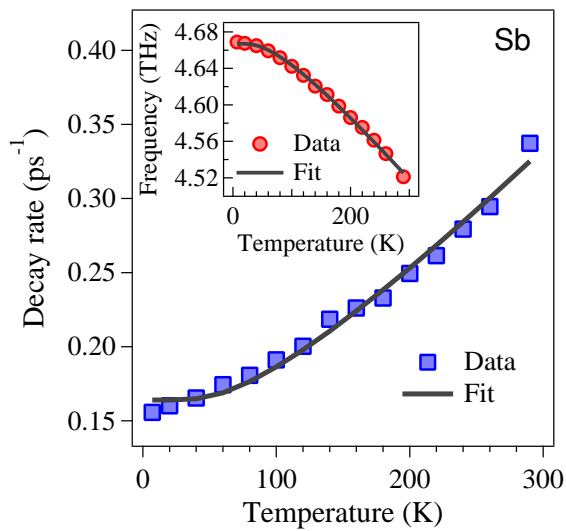}
\caption{Decay rate of the $A_{1g}$ mode together with the frequency (inset) as a function of the lattice temperature. The solid line represents the fit to the data 
using Eq. (2) with $\Gamma_{0}$ = 0.1 ps$^{-1}$ and $\Gamma$ = 0.06 ps$^{-1}$. In the inset the frequency of $A_{1g}$ mode is fit to the model given by Eqs. (3) and (4), which is represented by the solid line. 
}
\label{Fig2}
\end{figure}
The decay rates of the $A_{1g}$ mode monotonically increases as the lattice temperature rises, which is comparable to the results in Bi films.\cite{Hase98} 
This behavior is attempt to fit by an anharmonic decay model \cite{Klemens66}, in which the optical phonon decays into two acoustic phonons with half the frequency of 
the optical mode ($\omega_{A_{1g}}/2$) and with opposite wavevectors \cite{Klemens66,Balkanski83,Hase98,Hase05}, 
\begin{eqnarray}
\Gamma_{A_{1g}} = \Gamma_{0} +\Gamma(1 + 2n_{A_{1g}/2}),
\end{eqnarray}
where $\Gamma_{0}$ is a background contribution due to impurity and defect scattering, $\Gamma$ is the anharmonic coefficient, 
$n_{A_{1g}/2}$ = $[\exp(\hbar\omega_{A_{1g}}/2k_{B}T) -1]^{-1}$ is the Bose-Einstein factor, and $k_{B}$ is the Boltzmann constant. 
The fitting of the time-domain data to Eq. (2) is shown in Fig. 2, where we obtain $\Gamma_{0}$ = 0.1 ps$^{-1}$ and $\Gamma$ = 0.06 ps$^{-1}$ for Sb. 
Thus, the temperature dependence of the decay rate of the coherent $A_{1g}$ mode is well described by Eq. (2), 
indicating that the dephasing of the coherent $A_{1g}$ mode is governed by the anharmonic phonon-phonon coupling. 
In the same manner, the frequency softening of the $A_{1g}$ mode is expressed by taking into account contributions from the thermal expansion and anharmonic coupling, both of which are related to the cubic anharmonic term,\cite{Balkanski83,Cusco07}
\begin{eqnarray}
\omega_{A_{1g}} = \omega_{0} + \Delta_{0}(T) + \Omega(1 +2n_{A_{1g}/2}),
\end{eqnarray}
where $\omega_{0}$ is the harmonic frequency at the lowest temperature limit, $\Omega$ is an anharmonic constant, and 
{$\Delta_{0}$($T$) represents the frequency shift due to the thermal expansion of the lattice, which is given by, \cite{Cusco07} 
\begin{eqnarray}
\Delta_{0}(T) = - \omega_{0}\gamma \int_{0}^{T}[\alpha_{//}(T^\prime) + 2\alpha_{\perp}(T^\prime)]dT^\prime,
\end{eqnarray}
where $\gamma$ is the Gr{\"u}neisen parameter, $\alpha_{//}$($T^\prime$) and $\alpha_{\perp}$($T^\prime$) are the linear thermal expansion coefficients along directions parallel and perpendicular to the $c$-axis, respectively. In order to calculate $\Delta_{0}$($T$), we took the experimental linear thermal expansion coefficients and the Gr{\"u}neisen parameter ($\gamma$ = 1) measured by White,\cite{White72} and thus model fitting based on Eqs. (3) and (4) was carried out. 
The fitting result of the time-domain data to Eqs. (3) and (4) is shown in the Fig. 2 inset, where we obtain $\omega_{0}$ = 4.75 THz and $\Omega$ = - 0.087 THz, the later of which is comparable to the result ($\Omega$ = - 0.056 THz) in Bi films.\cite{Hase98} It should be mentioned here that the contribution from the thermal expansion to the frequency softening becomes effective at $T$ $\geq$ 100 K because the linear thermal expansion coefficients are rather small at $T$ $\leq$ 100 K, while they become larger and nearly constant at $T$ $\geq$ 100 K.\cite{White72}
\begin{figure}
\includegraphics[width=7.8cm]{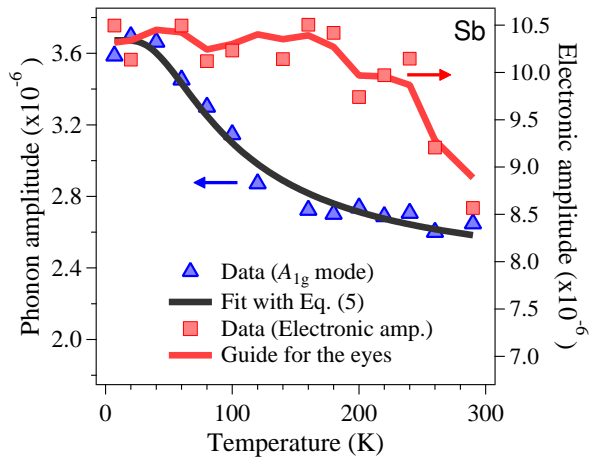}
\caption{
The amplitude of the coherent $A_{1g}$ mode ($A$) and the electronic component ($B$) as a function of the lattice temperature. 
The black solid curve represents the fit to the data using Eq. (5), using a fitting parameter of $A_{0}$ = 1.8$\times$10$^{-6}$ and a temperature independent background of 1.9$\times$10$^{-6}$. 
The red solid line represents the guide for the eyes. 
}
\label{Fig3}
\end{figure}
As shown in Fig. 3, the amplitude of the coherent $A_{1g}$ mode increases with decreasing the temperature. The ratio of the amplitude of the coherent $A_{1g}$ 
mode obtained at 7 K to that obtained at 290 K is $\approx$1.3. A similar temperature dependence of the coherent phonon amplitude was observed for the $E_{g}$ mode in a single crystal Sb by Ishioka {\it et al.}, while they observed rather flat 
behavior for the $A_{1g}$ mode.\cite{Ishioka08} 
The difference in the amplitude of the coherent $A_{1g}$ mode between the two measurements might partly come from the different temperature dependence of the electric component, since the amplitude of the $A_{1g}$ mode is proportional to the photoexcited carrier density, $A = \kappa B$, where $\kappa$ is an electron-phonon coupling constant, under the DECP model.\cite{Zeiger92} In fact, the one observed the increase of the electronic transient between 100 and 200 K by the isotropic reflectivity measurements,\cite{Ishioka08} while we observed rather decrease of the electric component with the temperature by the isotropic reflectivity measurements (see Fig. 3), as has been observed also in simple metals.\cite{Hase05} 
We fitted the temperature data of the phonon amplitude in Fig. 3 to the expected temperature dependence of the peaked intensity of spontaneous Raman scattering assuming a Lorentzian line shape with the linewidth controlled by anharmonic decay.\cite{Misochko06} By using the well-known property of a Lorentzian, the temperature dependence of peak intensity of the Raman-active $A_{1g}$ mode is proportional to; \cite{Misochko06,Kamaraju10}
\begin{eqnarray}
A = A_{0}\frac{n_{A_{1g}}+ 1}{2 n_{A_{1g}/2}+ 1},
\end{eqnarray}
where $A_{0}$ is a constant, $n_{A_{1g}}$ = $[\exp(\hbar\omega_{A_{1g}}/k_{B}T) -1]^{-1}$  is the Bose-Einstein factor, and the denominator of Eq. (5) represents the linewidth controlled by the anharmonic decay [see Eq. (2)]. 
The agreement between the experimental data and the Raman cross-section in Fig. 3 is very good for the temperature range of 7 K $\leq$ T $\leq$ 290 K. 
The good agreement of the experimental phonon amplitude with Eq. (5) found in Fig. 3 suggests the RSFL mechanism would predominate the generation of the coherent $A_{1g}$ mode over the DECP mechanism. 
This interpretation is further supported by the fact that the temperature dependence of the electronic amplitude ($B$) exhibits a significantly different nature from that of the coherent $A_{1g}$ mode ($A$) as shown in Fig. 3, and consequently the DECP model ($A = \kappa B$) cannot sorely account for the observed results.

\section{Discussion}
It is to be noted that in case of coherent phonons of Bismuth, the $E_{g}$ mode was also observed with $\approx$1/10 Fourier transformed (FT) spectral intensity at low temperatures even with the isotropic detection scheme.\cite{Ishioka06} In Sb, there seems to be no $E_{g}$ component at 7 K in Fig. 1, which was also confirmed by the FT spectra (not shown). A possible reason why the difference in the appearance of the $E_{g}$ mode between these two different atoms was found is polarization dependence of Raman tensor for the $E_{g}$ mode. In order to observe the $E_{g}$ mode by the isotropic detection one would need to use highly-oriented single crystal sample and to precisely rotate the pump-polarization in $x-y$ plane to match the appropriate crystal axis,\cite{Ishioka06} although the study of pump-polarization dependence of the $E_{g}$ mode using such a highly-oriented single crystal is beyond the scope of this paper. 

Hereafter we rather focus on the temperature dependence of the amplitude of the coherent phonon, referring other metallic system, since it will give a new insight into the physics of the coherent phonons in a wide range of the materials. 
A strong temperature dependence of the coherent optical phonon was observed in the simple metal Zn, in which the temperature dependence was examined in terms of the quasiparticle density.\cite{Hase05} 
Fig. 4 shows the comparison of the fitting models for the temperature dependence of the amplitude of the coherent optical phonon observed in Zn.\cite{Hase05} 

\begin{figure}
\includegraphics[width=7.5cm]{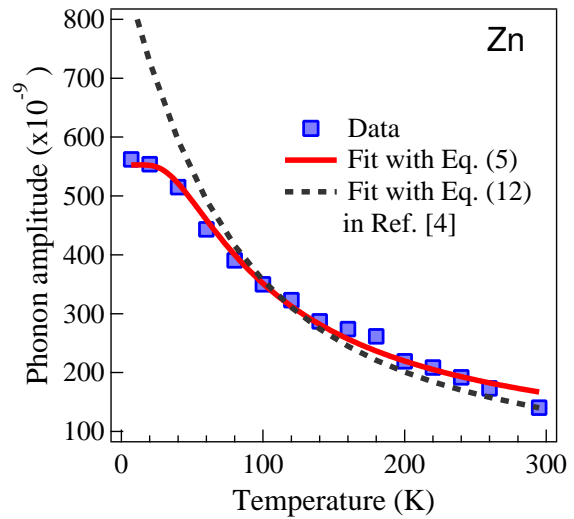}
\caption{The amplitude of the coherent $E_{g}$ mode in Zn as a function of the lattice temperature (the data were reproduced from Ref. [4]). The red solid line represents the fit to the data using Eq. (5), while the dashed line is the fit with Eq. (12) in Ref. [4]. 
}
\label{Fig4}
\end{figure}

It is found that the current model, Eq. (5) in the present study, rather well describes the overall data in Fig. 4, although the previous model based on the density of the quasiparticles also could fit the data at T $\geq$ 80 K while some deviation was found at lower temperature than 80 K.\cite{Hase05}  
Consequently we conclude that the overall behavior of the coherent phonon amplitude as a function of the lattice temperature is well modeled by Eq. (5) in a wide range of metallic systems (Bi, Sb, and Zn) or in the wide range of the Debye temperature, $\Theta_{D}$ $\approx$ 120 K for Bi, $\Theta_{D}$ $\approx$ 210 K for Sb and $\Theta_{D}$ $\approx$ 330 K for Zn.\cite{Kittel86} 
Note that a plausible reason why the deviation was observed at T $\leq$ 80 K for Zn in Fig. 4 between the model in Ref. [4] and the data is a classical nature of the simple model in Ref. [4], or $A \propto T_{e}^{\prime} - T_{e}$, where $T_{e}^{\prime}$ and $T_{e}$ are the final and initial electron temperature, respectively. The current model, Eq. (5), is rather based on quantum effect described by the Bose-Einstein factor. 

\section{Conclusion}
In conclusion, temperature dependence of the dynamics of coherent optical phonons in antimony single crystal has been studied by a femtosecond pump-probe technique at various lattice temperatures. 
The agreement of the decay rate and the frequency of the coherent $A_{1g}$ mode with the anharmonic model indicates that the dephasing of coherent phonons in antimony is 
dominated by anharmonic decay (energy relaxation), and that the frequency softening is due to cubic term of the anharmonicity, 
in which thermal expansion and anharmonic phonon-phonon coupling play important roles.
The coherent phonon amplitude of the $A_{1g}$ mode shows the decrease with the lattice temperature, 
which is well modeled by the use of the peaked intensity of spontaneous Raman scattering, and which also reproduces the amplitude of the coherent $E_{g}$ mode in Zn satisfactory. 
Thus, in a wide range of metallic systems anharmonic lattice effect dominates the temperature dependence of the decay rate, the frequency, and the amplitude of the photo-excited coherent optical phonons.

\begin{acknowledgments}
This work was supported in part by a Grant-in-Aid for the Scientific Research from the Ministry of Education, Culture, Sports, Science, and Technology of Japan under grant KAKENHI-15740188.
\end{acknowledgments}


\end{document}